\documentclass[pra,aps,twocolumn,superscriptaddress,longbibliography]{revtex4-1}
\usepackage[colorlinks=true, citecolor=red, urlcolor=blue ]{hyperref}
\usepackage{graphicx}
\usepackage{bm}
\usepackage{amsmath,amsfonts}
\usepackage{amsthm}
\usepackage{hyperref}
\usepackage{xcolor}
\hypersetup{
    urlcolor=blue,
    citecolor=blue
           }

\usepackage{braket}
\theoremstyle{plain}
\usepackage{color}
\usepackage{amssymb}
\usepackage{amsthm}
\usepackage{amsfonts}
\usepackage{float}
\usepackage{tabularx}
\usepackage{graphicx}

\usepackage{mathtools}
\usepackage{esvect}
\usepackage{wrapfig}
\usepackage{amsthm}
\usepackage{verbatim}
\usepackage{bbm}
\usepackage[normalem]{ulem}

\usepackage{enumitem}
\usepackage{fmtcount}
\usepackage{booktabs}
\usepackage{csquotes}
\usepackage{epsfig}

\usepackage{tabularx}
\usepackage{graphicx}
\usepackage{amsmath}
\usepackage{braket}
\usepackage{latexsym}
\usepackage{bm}
\usepackage{graphics,epstopdf}
\usepackage{enumitem}
\usepackage{fmtcount}
\usepackage{booktabs}
\usepackage{csquotes}
\usepackage{epsfig}

\theoremstyle{plain}

\def\bea{\begin{eqnarray}}
\def\eea{\end{eqnarray}}
\def\ba{\begin{array}}
\def\ea{\end{array}}

\def\beq{\begin{equation}}
\def\eeq{\end{equation}}

\usepackage[normalem]{ulem}
\usepackage{float}
\usepackage{graphicx}  
\usepackage{dcolumn}          
\usepackage{amssymb}
\usepackage{appendix}
\usepackage{physics}   
\usepackage{mathtools}
\usepackage{esvect}
\usepackage{wrapfig}
\usepackage{amsthm}
\usepackage{verbatim}
\usepackage{bbm}

\usepackage[mathscr]{euscript}
\def\Tr{\operatorname{Tr}}

\def\({\left(}
\def\){\right)}
\def\[{\left[}
\def\]{\right]}

\begin{document}
\title{\textbf {Harnessing energy extracted from heat engines to charge  quantum batteries}}
\author{Debarupa Saha}
\affiliation{Harish-Chandra Research Institute,  A CI of Homi Bhabha National Institute, Chhatnag Road, Jhunsi, Prayagraj  211019, India}
\author{Aparajita Bhattacharyya}
\affiliation{Harish-Chandra Research Institute,  A CI of Homi Bhabha National Institute, Chhatnag Road, Jhunsi, Prayagraj  211019, India}
\author{Kornikar Sen}
\affiliation{Harish-Chandra Research Institute,  A CI of Homi Bhabha National Institute, Chhatnag Road, Jhunsi, Prayagraj  211019, India}
\author{Ujjwal Sen}
\affiliation{Harish-Chandra Research Institute,  A CI of Homi Bhabha National Institute, Chhatnag Road, Jhunsi, Prayagraj  211019, India}
\begin{abstract}
We explore the performance of three- and two-stroke heat engines with a qutrit working substance in charging two-level quantum batteries. We first classify the heat engines into two groups depending on their working methods. The first type of heat engine, the sequential engine, evolves through three distinct strokes, viz., heat, work, and cold strokes. In the second kind of engine, a simultaneous engine, all the three events are made to occur simultaneously in one stroke, followed by an additional stroke to thermalize the working substance, i.e., the qutrit with a cold bath. We further categorize these two types of engines into two classes depending on the type of interaction between the working substance and the baths or the battery, viz., out-and-out engines, where the system bath interactions can invoke population transitions between any two energy levels of the qutrit, and fragmented engines, where only selective transition is materialized. Considering these four types of heat engines, we analyze the work done by the working substance, the percentage of charge accumulated by the quantum battery, and the efficiency of the engine. By drawing a comparison between the charging schemes, we  find that the sequential out-and-out heat engines are most advantageous, providing unit efficiency and transferring the most energy to the quantum battery, in the optimal case. The ranking of the benefits obtained from the other three engines depends on the quantity of interest.

\end{abstract}
\maketitle
\section{Introduction}

A classical heat engine is a device that converts heat energy into mechanical energy. The performance of quantum analogues of classical heat engines is currently an active area of research due to its potentially broad applicability. To the best of our knowledge, the first study on quantum heat engines was conducted by Scovil and Schulz-DuBois in 1959~\cite{pore1}. Later, Alicki~\cite{pore2} and Kosloff~\cite{pore3} extended the research by incorporating open-system formalism. The main motivation for studying quantum heat engines is to achieve improved performance by harnessing the quantum properties of the working substance. In this pursuit, the roles of quantum coherence~\cite{C2,C5,C6,E6,C3,C1,C11,C7,C10,C8,C9}, quantum correlations~\cite{co3,coi3,co1,co2,co4}, and interactions among the constituents~\cite{coi3,I3,pore4,I2,pore5,pore6,I1,I4,pore7} of the working body in the operation of thermal engines are being widely explored. Various fundamental aspects relevant to the study of heat engines include topics such as resource-theoretic limits on work extraction~\cite{pore8,pore9} and geometric approaches to thermodynamics~\cite{pore10,pore11,pore12}. Understanding the impact of thermal fluctuations on the performance of nano-engines constitutes another dynamic area of research in this direction~\cite{fl1,fl2}.

Refs.~\cite{E1,E4,E5,E3,E2} contain experimental realizations of quantum heat engines. The working substances used in these experimental setups include nitrogen vacancy centers~\cite{E6} and superconducting circuits~\cite{E9,E8,E7}. Scientists have also explored the mechanisms of heat engines with quantum dots~\cite{E10} and cold bosonic atoms~\cite{E11}. Heat engines have been extensively studied in various contexts, including optomechanical contexts~\cite{OM}, nanomechanical contexts~\cite{E12,E13}, and electromechanical settings~\cite{E14}.

Physical quantities used to assess the performance of a heat engine include the power generated per cycle and the engine's efficiency. Thermodynamic trade-offs among efficiency, power, and stability in quantum engines have also been established~\cite{ef2,TR1,TR2,TR3}. The enhancement of power in non-Markovian regimes is discussed in Refs. ~\cite{NP1,NP2,NP3}. Additionally, there are articles, such as Refs.~\cite{impw}, and \cite{bcl0,bcl}, where the concept of imperfect work has been introduced, along with examples where engine efficiency surpasses the Carnot efficiency.

While the most commonly studied engines consist of four distinct strokes, heat engines can also be modeled using three or even only two strokes~\cite{21,31,32,33,23,24,26,25,CB1}. The three-stroke engines are often referred to as minimally coupled engines~\cite{32,33}. These three strokes comprise two sequential strokes in which the working substance interacts with cold and hot baths, respectively, to release and absorb heat, and a third stroke in which the substance performs work. In addition to Markovian evolutions, three-stroke engines have also been studied in the presence of non-Markovian interactions~\cite{NP3}.

Two-stroke engines often consist of bipartite working substances~\cite{21}. Typically, in the first stroke, each of the parties interacts with their respective baths, while in the second stroke, the baths are disconnected, and the parties are coupled to each other. There are also other notions of two-stroke engines. For instance, in Refs.~\cite{23,24,26,25}, the authors dealt with stroboscopic two-stroke quantum heat engines based on a collision model. Various unconventional heat engines have also been introduced, such as measurement-based heat engines where the interactions of the working substances with the thermal baths have been replaced by measurements performed on the working substances~\cite{2M,22}, or continuous or simultaneous engines~\cite{21,HC}, where all three processes, i.e., heat absorption, work done, and relinquishment of heat, occur simultaneously.

In this article, we investigate the performance of sequential (three-stroke) and simultaneous (two-stroke) heat engines for charging quantum batteries. Both types of engines consist of a qutrit working substance, a two-level quantum battery that we want to charge, and two bosonic baths prepared at different temperatures. The three strokes of the sequential engines involve interaction between the working substance and the 1) hot bath, in the first stroke, 2) quantum battery, in the second stroke, and 3) cold bath, in the third stroke.

In the sequential three-stroke engines, these three events occur consecutively. In the simultaneous engine, all three events are merged to occur together in a single stroke. To restore the initial state of the working substance in the simultaneous engine, we introduce a second stroke where the working substance interacts with the cold bath only and reaches equilibrium with it.

Further, the sequential and simultaneous engines can be classified into two categories: the out-and-out engines and the fragmented engines, depending on the types of interaction the working substances undergo with the baths or the quantum batteries. In the out-and-out engines, the baths can impose a transition between any two energy levels of the qutrit working substance, while in the fragmented engines, population transition between selective energy levels is permitted.

We quantify the performance of these engines by evaluating the work done by the working substances, the percentage of charge gained (PCG) by the batteries, and the engine's efficiency. Upon comparing the engines, we find that the sequential out-and-out engines are the most advantageous among the four. The ranking of the other three engines depends on the specific quantity of interest. For instance, the sequential fragmented engines perform better in terms of work done compared to the simultaneous fragmented engines, while the latter are more efficient than the former.

The charging of quantum batteries using thermal engines has been previously discussed in Refs.~\cite{CB1,CB2,32,33,HC}. Though the engines discussed in Refs.~\cite{32,33} have three sequential strokes, there is a notable distinction between our approach and the works in Refs.~\cite{32,33}. In both of those works~\cite{32,33}, the authors aimed to transfer ergotropy from the working substance to the battery. In contrast, our sequential out-and-out engine, as considered here, initiates the interaction between the battery and the working substance only after the working substance has reached equilibrium with the hot bath. At this point, the working substance accumulates a passive state with zero ergotropy. Therefore, the energy being transferred to the quantum battery is a portion of the passive energy of the working substance. Ref.~\cite{HC} deals with a simultaneous engine, but their model significantly differs from our simultaneous engine's model. They considered qubit baths and described the interactions between the working substance and the baths or quantum battery using a collisional-type model. Additionally, our approach differs in that we examine the heat engines using specific models and focus more on quantitative behavior. Similarly, Refs.~\cite{CB1} and \cite{CB2} also fundamentally differ from the work presented in this paper.

The remainder of the paper is structured as follows: In Sec.~\ref{tools}, we elaborate on the essential equipment required for modeling the engines. Secs.~\ref{seq} and~\ref{sim} delve into the discussion of the out-and-out and fragmented sequential engines, and the performance of the out-and-out and fragmented simultaneous engines, respectively.
Sec.~\ref{seclast} offers a brief comparison between the engines. Lastly, in Sec.~\ref{Con}, we present our concluding remarks.

\section{UNDERLYING TOOLS}
\label{tools}
Let us discuss the details of the structures of the heat engines. We consider a qutrit system, ${S}$, as the working substance of our engines and hot and cold bosonic baths, say $H$ and $C$, respectively. The motive of these heat engines is to charge two-level quantum batteries, ${B}$. 
Without loss of generality, we can take the local Hamiltonian of the qutrit of the form
\begin{equation}
\label{He}
    H_{{S}}=
\begin{bmatrix}
  \frac{A}{2} & 0 & 0\\
  0 & -\frac{A}{2}& 0\\
  0 & 0 & 0
\end{bmatrix},
\end{equation}
with the only consideration that the energy gap between the first and second energy levels is equal to the same between the second and third energy levels. Let the eigenstates of $H_S$ be $\{\ket{\tilde{0}},\ket{\tilde{1}},\ket{\tilde{2}}\}$, where the corresponding eigenvalues are $\{-A/2,0,A/2\}$. 

The Hamiltonian of a Bosonic bath, $X$, can be taken to be
\begin{equation}
H_{{X}}=\hbar\int_{0}^{\omega_{X}^c}a_{\omega}^{\dagger}a_{\omega}d\omega,
\label{HX}
\end{equation}
where $a^\dagger_\omega$ ($a_\omega$) denotes the creation (annihilation) operator which creates (annihilates) a particle with angular frequency $\omega$. $a_\omega^\dagger$ and $a_\omega$ satisfy the bosonic commutation relation, viz. $\left[a_\omega,a_{\omega '}^\dagger\right]=\delta_{\omega\omega'}$. $\omega_X^c$ and $\hbar$ represent, respectively, the frequency of the highest mode of the bath, $X$, and Planck's constant. We can choose $X$ to be $H$ or $C$ depending on if the interacting bath is hot or cold, respectively. 

The energy of the quantum battery can be defined in terms of the following Hamiltonian:
\begin{equation}
\label{HB}
    H_{{B}}=\frac{A}{4}\sigma_{z},
\end{equation}
where $\sigma_z$ is a Pauli matrix. We will denote the other two Pauli matrices by $\sigma_x$ and $\sigma_y$. The eigenvalues and eigenvectors of $H_B$ are $\{-A/4,A/4\}$ and $\{\ket{0},\ket{1}\}$, respectively.
To charge the quantum battery, the qutrit interacts with a hot bath, a quantum battery, and a cold bath. Depending on whether these three processes take place one by one or all at once, we characterize heat engines in two parts, i.e., the ``sequential heat engines" and ``simultaneous heat engines". Heat engines can further be categorized into two groups, i.e., ``out-and-out heat engines" and ``fragmented heat engines," based on the type of interaction the working material undergoes with the baths and quantum batteries.

Whatever be the type of the engine, we always consider the interaction of the working substance with the baths weak enough for the evolution to be Markovian.
Evolution of any system, $M$, weakly coupled to a bath can be determined using the Gorini–Kossakowski–Sudarshan–Lindblad (GKSL) master equation~\cite{gk1,gk2,gk3}, 
\begin{equation}
\label{gksl}
    \frac{d\rho}{dt}=-i/\hbar[H_{M},\rho]+\mathcal{L(\rho)},
\end{equation}
where $\rho$ is the state of $M$ and $H_M$ is the system's local Hamiltonian.
The first term of the master equation represents the evolution of the system dictated by its own Hamiltonian, and $\mathcal{L(\rho)}$ is the dissipative term that occurred due to the presence of the bath. Explicit expression of $\mathcal{L}(\rho)$ is provided below:
\begin{widetext}
\begin{eqnarray}
    \mathcal{L}(\rho)&=&\sum_{i,k,l}\Gamma_{kl}(\omega_i)\Big[A_l(\omega_i)\rho A_k^\dagger(\omega_i)-\frac{1}{2}\{A_k^\dagger(\omega_i)A_l(\omega_i),\rho \}\Big]+\sum_{i,k,l}\Gamma_{kl}(-\omega_i)\Big[A_l^\dagger(\omega_i)\rho A_k(\omega_i)-\frac{1}{2}\{A_k(\omega_i)A_l^\dagger(\omega_i),\rho \}\Big]\nonumber\\
    &=&\mathcal{L}_+(\rho)+\mathcal{L}_-(\rho),\label{eq1}
\end{eqnarray}
\end{widetext}
where $\omega_i$ is the energy difference between $i$th pair of energy levels between which transition is possible due to the bath and $\Gamma_{kl}(\omega_i)$ and $A_k(\omega_i)$ are the corresponding transition rate and Lindblad operator. $\omega_i\neq \omega_j$ for all $i\neq j$. The explicit forms of $\omega_i$, $\Gamma_{kl}(\omega_i)$, and $A_k(\omega_i)$ can vary from engine to engine depending on the properties of the interaction between the system and the bath. We will use Eq.~\eqref{eq1} to determine the evolved states of the working substances in the presence of a bath.

The physical quantities through which the performance of the heat engines is reflected are 
\begin{itemize}
    \item the total amount of heat consumed by the working substance from the baths ($Q_{T}^P$), and
    \item the energy transferred to the quantum battery ($W_B^P$).
\end{itemize}
These quantities can further be used to determine the percentage of charge gained (PCG) by the battery, $C^P=\frac{\text{energy absorbed by $B$}}{\text{energy gap of $B$}}\times 100\%=\frac{2W_B^P}{A}\times 100\%$, and the efficiency, $\eta^{P}=W_B^P/Q_{T}^P$, of the engine. Usually, the working substance operating in the heat engines absorbs heat from the hotter bath and releases heat to the colder one. Therefore, in those cases $Q_{T}^P$
becomes equivalent to the heat taken from the hot bath, $Q_H^P$. In this work, we will encounter situations where working substances will be found to be absorbing heat from both baths [in Secs. \ref{ok} and \ref{ok2}]. Therefore, we consider such a definition of efficiency where the total absorbed heat is taken into consideration instead of the contribution of the hot bath alone. 

We use the notation $P=1O,~1F,~2O,$ and $2F$ to denote respectively out-and-out sequential engines, fragmented sequential engines, out-and-out simultaneous engines, and fragmented simultaneous engines.

In the remaining part of the paper, we will analyze the effectiveness of the heat engines through the characteristic quantities $W_B^P$, $C^P$, and $\eta^P$. There may exist particular engines for which $\eta^{P}$ is much high but $W_B^{P}$ is negligible. Therefore, to examine the nature of a profitable engine, we will first maximize the amount of energy that can be transferred to the quantum battery, i.e., $W_B^{P}$, over the set of free parameters which specifies the engine and then determine $C^P$ and $\eta^P$ for that optimal parameters. Finally, we will compare the optimal performances of the four types of engines by inspecting the behaviors of $W_B^P$, $C^P$, and $\eta^P$.

  \begin{figure*}
\centering
\includegraphics[scale=0.40]{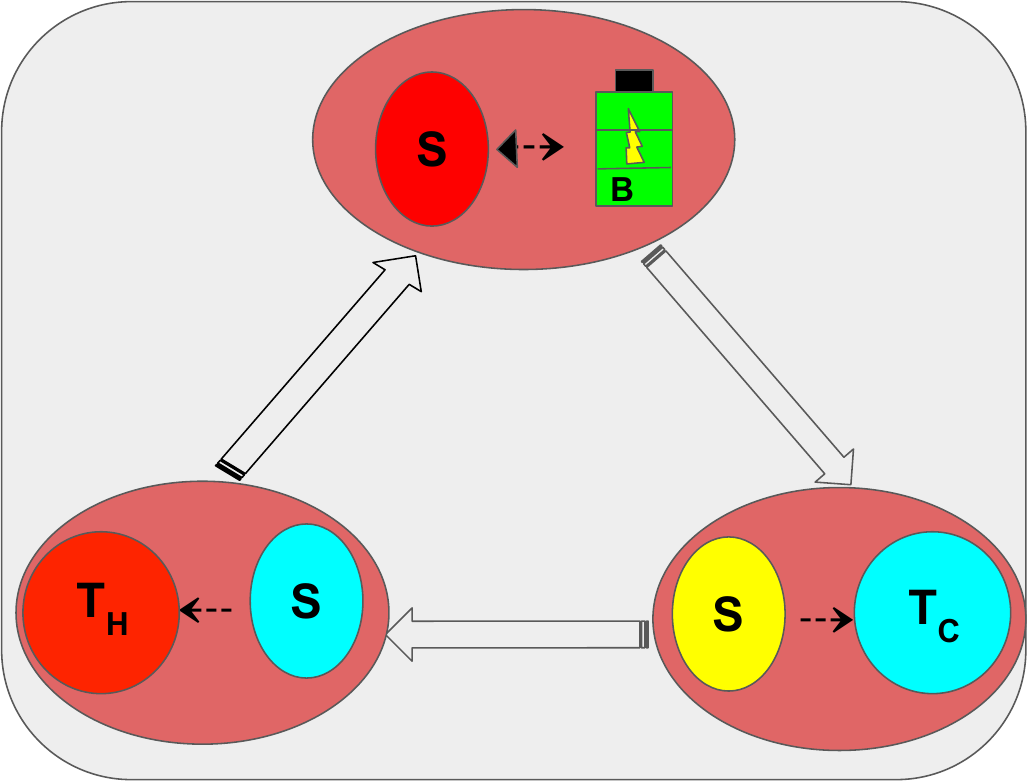}
 \hspace{1.2cm}
\includegraphics[scale=0.50]{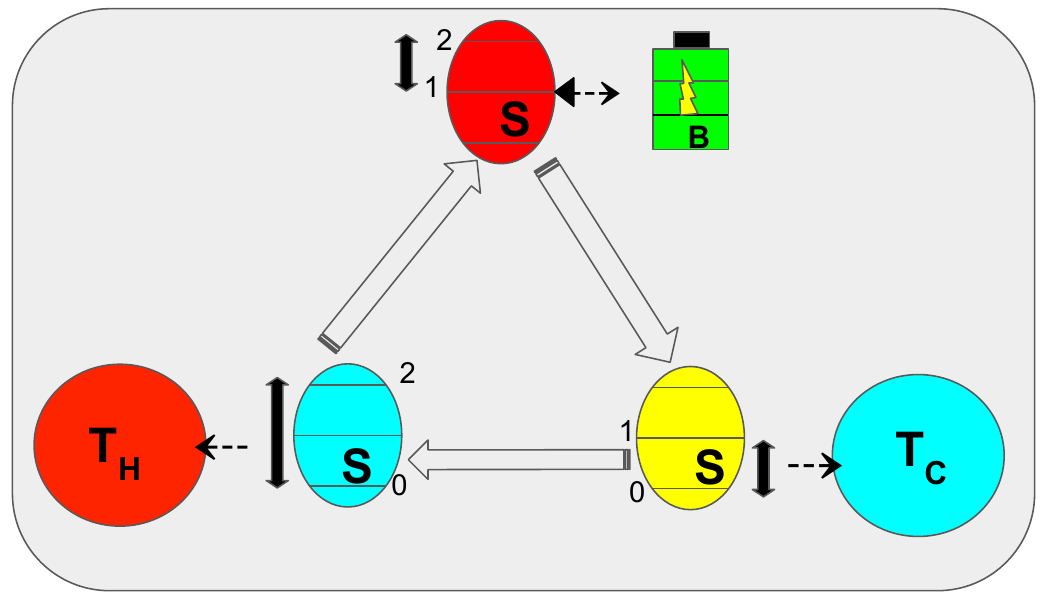}
	\caption{Diagrammatic representation of sequential engines: The left and right panels demonstrate the working procedures of the out-and-out and fragmented sequential engines, respectively. The circles represent the baths, and the system is depicted using red, yellow, or sky-blue elliptical shapes. $T_H$ and $T_C$ are, respectively, the temperatures of the hot and cold baths. Both engines consist of three distinct strokes, during each of which the working substance ($S$) interacts with either the hot bath, cold bath, or quantum battery ($B$). The large, colorless arrows indicate the direction of time, while the dashed arrows show the direction of information flow in each stroke. Since the interactions between the working substance and the baths are considered to be Markovian in nature, information flows only from the system to the baths. However, when $S$ and $B$ interact with each other, information can flow in both directions. In the case of the out-and-out engine, population transition due to interaction is possible between any two energy levels of $S$. We indicate such interactions using brown, large ellipses containing the system and one bath or the quantum battery. Population transitions between specific energy levels of $S$, which occur in the fragmented engines, are indicated using black solid arrows. }
	\label{Pic1}
\end{figure*} 

\section{Sequential engines}
\label{seq}
As the name ``sequential engines" implies, in this type of engine, the three events—interaction with the hot bath, quantum battery, and cold bath—happen one by one in three distinct strokes. In the following, we inspect the three sequential strokes of the engines individually.
\begin{enumerate}
   \item{ \textbf{Heat Stroke:}} The cycle begins with the working substance, $S$, being in thermal equilibrium with the cold bath having a fixed temperature, say $T_C$. Thus, the initial state of $S$ is a thermal state of the form $$\rho_{C}^{1p}=\frac{e^\frac{-H_{S}}{k_BT_{C}}}{\Tr\left(e^\frac{-H_{S}}{k_BT_{C}}\right)},$$ where $k_B$ is the Boltzmann constant. In the first stroke, $S$ is decoupled from the cold bath, $C$, and allowed to weakly interact with a hot bath, $H$, of temperature $T_H>T_C$ for a large amount of time so that it can reach its steady state, $\rho_{H}^{1p}$. This is the ``heat stroke", where heat gets extracted from the hot bath to the working material, $S$. The total amount of heat, $Q_H^{1p}$, taken from the hot bath through the evolution is given by the change in internal energy  of $S$, i.e.,
\begin{equation*}
    Q_{H}^{1p}=\Tr[H_{S}(\rho^{1p}_{H}-\rho^{1p}_{C})].
\end{equation*}
\item{\textbf{Work Stroke:}}
 In this stroke, a portion of the energy absorbed from the heat bath, $H$, is transferred to the battery, $B$, through the working substance, $S$. We refer to this as the ``work stroke".
 To accomplish the job, we decouple $S$ from $H$ and bring the battery, $B$. The composite initial state of $S$ and $B$ is a product state of the form
\begin{equation*}
\rho^{1p}_{SB}=\rho^{1p}_{H}\otimes\rho^{1p}_{B}.
\end{equation*}
Here $\rho^{1p}_{B}=\ket{0}\bra{0}$ represents the ground state of the battery. A particular interaction, $H^{1p}_{S{B}}$, is switched on between $S$ and ${B}$, which may transfer energy from the former to the latter or vice versa but will not be able to increase the overall energy of the entire system consisting of $S$ and $B$.  The total Hamiltonian of the composite system, ${S{B}}$, is of the form
\begin{equation*}
   \label{htotal} H_{\text{WS}}^{1p}=H_{S}\otimes\mathcal{I}_{2}+\mathcal{I}_{3}\otimes H_{{B}}+H_{S{B}}^{1p},
\end{equation*}
where $\mathcal{I}_{2/3}$ is the identity operator acting on two/three-dimensional Hilbert space. 
To ensure that the interaction Hamiltonian does not incorporate any additional energy into ${S{B}}$ during the evolution, it is necessary that $H_{{S{B}}}^{1p}$ commute with $H_{\text{WS}}^{1p}$, i.e., $[H^{1p}_{{S{B}}},H^{1p}_{\text{WS}}]=0$. 

After the time interval, $t_1$, the interaction between $S$ and $B$ is switched off. Hence, the final system-battery state at the end of this stroke is
\begin{equation}
    \Bar{\rho}^{1p}_{SB}=U\rho^{1p}_{SB}U^\dagger,\nonumber
\end{equation}
where $U=\exp(-iH^{1p}_{\text{WB}}t_1/\hbar)$. So the local state of the quantum battery or the working substances after the unitary evolution is given by
\begin{equation}
    \Bar{\rho}^{1p}_{B/S}=\Tr_{S/B}(\Bar{\rho}^{1p}_{SB}),\nonumber
\end{equation}
where $\Tr_{S/B}$ represents tracing out the system/battery.
In this process, the work done, $W_B^{1p}$, by $S$ to charge the quantum battery, $B$, is equal to the change in internal energy of the quantum battery, which is given by
\begin{equation}
\label{work}
    W_{B}^{1p}=\Tr[H_{B}(\Bar{\rho}^{1p}_{B}-\rho_{B})].\nonumber
\end{equation}

\item{\textbf{Cold Stroke:}} This is the last stroke of the three-stroke cycle. In this part, $S$ is weakly coupled to the cold bath of temperature $T_C$ for a large amount of time till it reaches a steady state, $\Bar{\rho}_{C}^{1p}$. The total amount of heat released to the cold bath in this process is 
\begin{equation}
    Q_{C}^{1p}=\Tr[H_{S}(\Bar{\rho}^{1p}_{S}-\Bar{\rho}^{1p}_{C})].\nonumber
\end{equation}
where $\theta(\cdot)$ is the Heavy side step function. Here $Q_C^{1p}<0$ indicates the case where $S$ absorbs heat from $C$. This type of situation is not encountered in classical heat engines; nevertheless, we will realize, at some moment, that there are certain temperatures for which $Q_C^{1p}$ of the considered quantum heat engine can become negative.

Once $W_B^{1p}$, $Q_{C}^{1p}$, and $Q_{H}^{1p}$ are obtained, the efficiency of the engines and PCG by the quantum battery can be calculated using the following expressions:
\begin{eqnarray*}
    \eta^{1p}&=&\frac{W_B^{1p}}{Q_T^{1p}}=\frac{W_B^{1p}}{Q_{H}^{1p}-Q_C^{1p}\theta\left(-Q_{C}^{1p}\right)}\text{ and }\\
   C^{1p} &=&\frac{2W_B^{1p}}{A}\times 100\%,
\end{eqnarray*}
\end{enumerate}

We categorize the sequential engines into two different classes depending on the nature of the interaction between $S$ and the baths, $X$, which we referred to as
\begin{enumerate}
    \item out-and-out three-stroke heat engines and
    \item fragmented three-stroke heat engines.
\end{enumerate}
To get a schematic understanding of the working method of these two types of engines, see Fig.~\ref{Pic1}. Detailed discussions about them are presented in the two following sub-sections. We will use the notations $p=O$ and $p=F$ to represent the out-and-out and fragmented engines, respectively. 

\subsection{ Out-and-out three-stroke heat engines}
\label{ok}
\begin{figure*}[t]
\centering
	\includegraphics[scale=0.59]{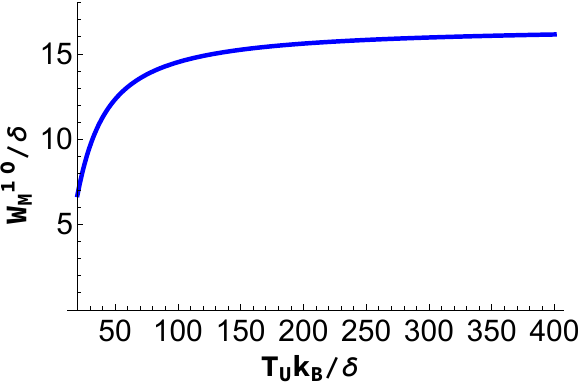}
	\includegraphics[scale=0.59]{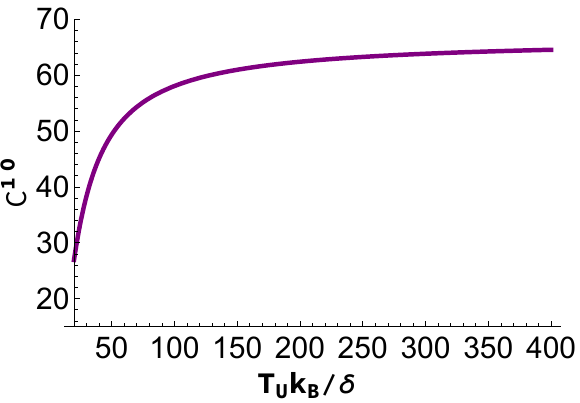}
	\includegraphics[scale=0.59]{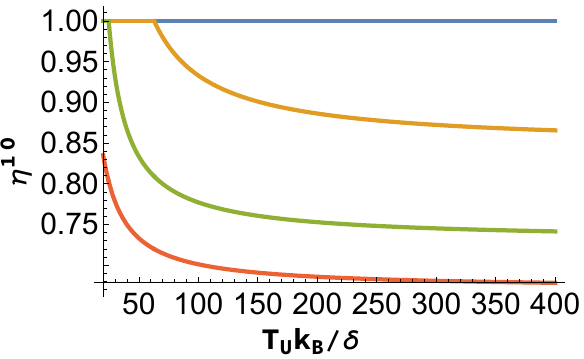}
	\caption{Unveiling the dependency of the three main figures of merit of a sequential out-and-out engine on the upper bound restricted to the temperature of the hot bath. We maximize the energy obtained by the battery from the working substance over the free parameters defining the out-and-out engine by putting the constraint that the hot bath's temperature should be less than $T_U$. In the vertical axes of the left panel, we plot the variation of this maximum energy transferred to the battery ($W_M^{1O}$) with respect to $T_U$, shown in horizontal axes. The middle and right panels show behavior of the percentage of gained charge $(C^{1O})$ and efficiency $(\eta^{1O})$ of the engine for the same parameters for which maximum work is obtained. $W_M^{1O}$ and $C^{1O}$ are independent of the temperature, $T_C$, of the cold bath whereas $\eta^{1O}$ depends on $T_C$. Thus the right panel contains four plots corresponding to four different cold bath temperatures, that are, 5$\delta/k_B$ (red curve), 10$\delta/k_B$ (green curve), 15$\delta/k_B$ (yellow curve), and 20$\delta/k_B$ (blue curve). All the axes of the plots are dimensionless.}
	\label{fig1}
\end{figure*} 

In this type of engine, we consider the interaction between the baths and the qutrit to be capable of inducing a population transition between any two energy levels of the working material.
 
Specifically, we consider the interaction between bath, $X$, and the working substance, $S$, to be described by the following Hamiltonian:
\begin{equation*}
\label{1}
\begin{split}
H^{1O}_{SX}&=\int_{0}^{\omega_{X}^c}h(\omega)\Big[\tilde{a}_{01}\left(a_{\omega}^X\right)^{\dagger } +\tilde{a}_{10}a_{\omega}^X+\tilde{a}_{02}\left(a_{\omega}^X\right)^{\dagger }  
\\&+\tilde{a}_{20}a_{\omega}^{X}
+\tilde{a}_{12}\left(a_{\omega}^X\right)^{\dagger } +\tilde{a}_{21}\left(a_{\omega}^X\right)^{\dagger }\Big]d\omega,
\end{split}
\end{equation*}
where $h(\omega)$ is the coupling constant and $\tilde{a}_{ij}=\ket{\tilde{i}}\bra{\tilde{j}}$. 
 
At the beginning of the heat stroke, the working substance's state is fixed at $\rho_C^{1O}=\frac{e^\frac{-H_{S}}{k_BT_{C}}}{\Tr\left(e^\frac{-H_{S}}{k_BT_{C}}\right)}$. Since the hot bath can lead to population transition between any two energy levels of $S$ in the heat stroke, at the end of that stroke the working material will reach the thermal state of temperature $T_H$, i.e., $\rho_H^{1O}=\frac{e^\frac{-H_{S}}{k_BT_{H}}}{\Tr\left(e^\frac{-H_{S}}{k_BT_{H}}\right)}$. Hence, the work stroke begins with a qutrit-battery state that is diagonal in the eigenbasis of the local Hamiltonian, $H_{S}\otimes\mathcal{I}_{2}+\mathcal{I}_{3}\otimes H_{B}$. 
Therefore, the only part of the Hamiltonian, $H_{\text{WS}}$, responsible for the evolution of $SB$ in the work stroke, is the interaction part, $H^{1O}_{SB}$. Therefore, we can use the unitary $U=\exp(-i(H^{1O}_{SB}t_1/\hbar))$ to evolve $\rho^{1O}_{SB}$ and get the final state, $\Bar{\rho}^{1O}_{SB}$. 
We take the interaction Hamiltonian $(H^{1O}_{{S{B}}})$ of the form
\begin{equation}
\label{Heb}
H^{1O}_{{S{B}}}=\hbar\omega_{sb}(J_{+}\otimes\sigma_{-}+J_{-}\otimes\sigma_{+}),
\end{equation}
where, $J_{+}=(\ket{\tilde{1}}\bra{\tilde{0}}+\ket{\tilde{2}}\bra{\tilde{1}})$, $J_{-} =J_+^\dagger$, $\sigma_+=\ket{1}\bra{0}$, and $\sigma_-=\sigma_+^\dagger$. 
We have checked that $[H_{SB}^{1O},H_{WS}^{1O}]=0$, which ensures that the evolution can only exchange energy between $S$ and $B$ and cannot increase or decrease the energy of the whole system, $SB$. Since the term $H^{1O}_{SB}/\omega_{sb}$ does not include any parameters, we can consider $U$ to be a function of a single unitless parameter, $\lambda=\omega_{sb}t_1$. In this work stroke, the quantum battery receives energy from the qutrit, after which the system is traced out and the qutrit is attached to the cold bath to begin the cold stroke. After a large amount of time, $S$ will again reach the initial thermal state, $\Bar{\rho}^{1O}_C=\rho_C^{1O}=\frac{e^\frac{-H_{S}}{k_BT_{C}}}{\Tr\left(e^\frac{-H_{S}}{k_BT_{C}}\right)}$ completing the cycle. 

Let us now discuss the characteristic quantities of the engine, i.e., $W_B^{1O}$, $C^{1O}$, and $\eta^{1O}$. In the case of this out-and-out sequential heat engine, the amount of heat extracted from the hot bath, $Q_H^{1O}$, and the energy transferred to the quantum battery, $W_B^{1O}$, are found to be
\begin{eqnarray}
    Q_{H}^{1O}&=\frac{A (2 \sinh (c-h)+\sinh (c)-\sinh (h))}{(2 \cosh (c)+1) (2 \cosh (h)+1)}~\text{and}\nonumber\\
    W_B^{1O}&=\frac{A\left(1+e^h\right)\left(1-\cos(2\lambda)\right)}{4\left(1+e^{h}+e^{2h}\right)},\label{eq10}
\end{eqnarray}
where $h=\frac{A}{2k_BT_H}$ and $c=\frac{A}{2k_BT_C}$. The heat absorbed by the cold bath, $Q_C^{1O}$, in the cold stroke can be easily calculated using the relation $Q_C^{1O}=Q_H^{1O}-W_B^{1O}$. 

The maximum energy that can be transferred to the quantum battery, optimized over $\lambda$, can be determined using Eq.~\eqref{eq10} and is given by 
\begin{equation}
    W_M^{1O}=\frac{A(1+e^h)}{2(1+e^h+e^{2h})}.\nonumber
\end{equation}
As one can notice from the above expression, $W_M^{1O}$ is a monotonically increasing function of both $A$ and $T_H$ which implies that the hotter the bath, the more charge the quantum battery receives. Therefore, if we maximize $W_m^{1O}$, by putting the restriction that the temperatures of the baths should be less than $T_U$, the resulting optimal temperature of the hot bath will be equal to $T_U$. Let us denote that maximum work, maximized over $T_H\leq T_U$, by $W_M^{1O}$.  

We introduce an energy unit, $\delta$, in terms of which all the quantities under consideration can be represented. To get a precise understanding of the nature of $W_M^{1O}$, in the left panel of Fig.~\ref{fig1}, we plot $W_M^{1O}/\delta$ with respect to $T_Uk_B/ \delta$ for $A=50\delta$. The behavior of PCG by the quantum battery (middle panel) is also presented in the same figure with respect to $T_Uk_B/ \delta$ taking $\lambda=\pi/2$, which maximizes the work done. 

The efficiency of the sequential out-and-out engines depends not only on $A$, $T_H$, and $\lambda$ but also on the temperature of the cold bath, $T_C$. Keeping $\lambda$ and $A$ fixed at, respectively, $\pi/2$ and $50\delta$, we plot $\eta^{1O}$ on the right panel of Fig.~\ref{fig1} for different cold bath temperatures, $T_C$. Readers might notice from the figure that there exists a wide range of $T_U$ for $T_C=10\delta/k_B$ and $15\delta/k_B$ for which the engine functions with unit efficiency. Actually, when $T_H$ is not much larger than $T_C$, the working substance starts absorbing heat in both the hot and cold strokes. Therefore, in this case, $Q_C^{1O}$ becomes negative, and, as a result, efficiency reaches unity. The reason behind this behavior is that when $T_C$ is sufficiently high, after providing energy to the quantum battery, the self-energy of $S$ becomes much smaller than that of the thermal state of $S$ with temperature $T_C$. Therefore, to reach equilibrium, $S$ absorbs heat from $C$ in the cold stroke. For smaller $T_C$, where $Q_C^{1O}> 0$, it is evident from the figure that though at first the work done increases with the hot bath's temperature, the efficiency decreases, both finally saturating to a constant value.

\subsection{Fragmented three-stroke engines}
Unlike the out-and-out three-stroke heat engine, in this fragmented three-stroke engine, the interactions between the baths and the qutrit are considered to be invoking an energy transition within selective pairs of the qutrit's energy levels. Specifically, the hot and cold baths can make transitions between $\ket{\tilde{0}}$ and $\ket{\tilde{2}}$ and $\ket{\tilde{0}}$ and $\ket{\tilde{1}}$, respectively. We want to explore the performance of such a heat engine where population transitions between certain pairs of energy levels of the qutrit are forbidden. The interaction Hamiltonian representing such a connection of $S$ with the hot bath, $H$, is given by
\begin{equation} H_{SH}^{1F}=\int_{0}^{\omega_{H}^c}h(\omega)\left(\tilde{a}_{02}\left(a_{\omega}^H\right)^{\dagger}
+\tilde{a}_{20}a_{\omega}^{H}
\right)d\omega.\nonumber
   \end{equation}
Since only one type of transition is possible, in this case, $\omega_i$ takes only one value, $\omega_H=A$, which corresponds to the transition $\ket{\tilde{0}}_{E}{\rightleftarrows} \ket{\tilde{2}}_{E}$. Therefore, the Lindblad operators are $A_{1}^{1F}(\omega_H)=J_{--}$ and $A_{2}^{1F}(\omega_H)=-iJ_{--}$, where $J_{--}=\ket{0}\bra{2}$ and $J_{++}=J_{--}^\dagger$. By evaluating $\Gamma_{kl}$, we obtain the forms of $\mathcal{L}_+$ and $\mathcal{L}_-$ defined in Eq.~\eqref{eq1}, which we denote as $\mathcal{L}_+^H$ and $\mathcal{L}_-^H$, respectively. The mathematical expressions of $\mathcal{L}_+^H$ and $\mathcal{L}_-^H$ are presented below:
\begin{eqnarray}
     \mathcal{L}_{+}^H=&
 \mathcal {J}(\omega_H)(1+n(\omega_H,T_H))\Big[A_1^{1F}(\omega_H)\rho \left(A_1^{1F}(\omega_H)\right)^\dagger\nonumber\\
&-\frac{1}{2}\Big\{\rho,\left(A_1^{1F}(\omega_H)\right)^\dagger A_1^{1F}(\omega_H)\Big\}\Big],\label{eq2}\\
\mathcal{L}_{-}^H=&\mathcal{J}(\omega_H)n(\omega_H,T_H)\Big[\left(A_1^{1F}(\omega_H)\right)^\dagger \rho A_1^{1F}(\omega_H)\nonumber\\
&- \frac{1}{2}\Big\{\rho,A_1^{1F}(\omega_H) \left(A_1^{1F}(\omega_H)\right)^\dagger\Big\}\Big]\label{eq3}.
\end{eqnarray}
Here $n(\omega_{H},T_H)$ is the average number of bosons in the bath of temperature, $T_H$, with frequency $\omega_H$, i.e., $n(\omega_{H},T_H)= \frac{1}{e^\frac{\hbar\omega_{H}}{k_BT_{H}}-1} $, and $\mathcal{J}(\omega_H)=\pi h^2(\omega_H)/2=\kappa \omega_H \exp(-\omega_H/\omega^c_H)$. Since $A_2^{1F}(\omega_H)$ can be expressed in terms of $A_1^{1F}(\omega_H)$, the former does not appear in the expressions of $\mathcal{L}^H_\pm$.

The interaction Hamiltonian characterizing the interaction between $S$ and $C$ is 
\begin{equation}
H_{SC}^{1F}=\int_{0}^{\omega_{C}^c}h(\omega)\left(a_{01}\left(a_{\omega}^C\right)^{\dagger }
+a_{10}a_{\omega}^{C}
\right)d\omega.\nonumber
\end{equation}
The transition gap in this case is $\omega_i=\omega_C=A/2$. The corresponding Lindblad operators are given by
 $A_{3}^{1F}(\omega_C)=\tilde{a}_{01}$ and $A_{4}^{1F}(\omega_C)=-i\tilde{a}_{01}$.  
The expressions of $\mathcal{L}_+$ and $\mathcal{L}_-$ involved in Eq.~\eqref{eq1} remain the same as $\mathcal{L}^H_+$ and $\mathcal{L}^H_-$ shown in Eqs.~\eqref{eq2} and \eqref{eq3}, respectively, with the only difference that $\omega_H$, $T_H$, and $A_1^{1F}$ will get replaced by $\omega_C$, $T_C$, and $A_3^{1F}$, respectively. We denote these dissipation terms as $\mathcal{L}_+^C$ and $\mathcal{L}_-^C$.
\begin{figure*}
\centering
	\includegraphics[scale=0.59]{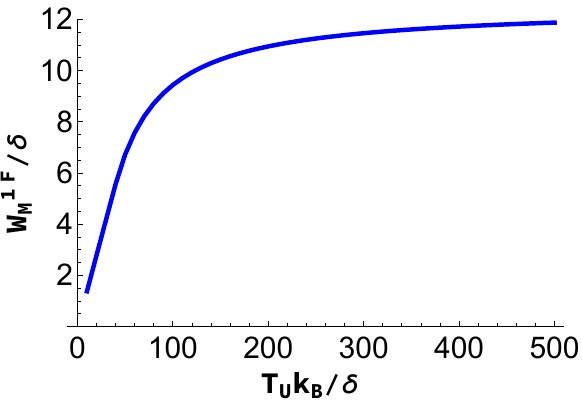}
	\includegraphics[scale=0.59]{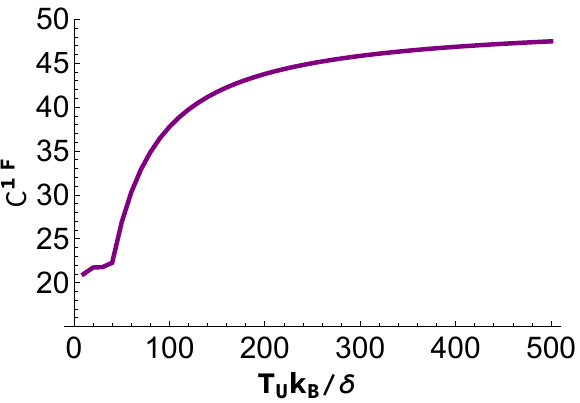}
	\includegraphics[scale=0.59]{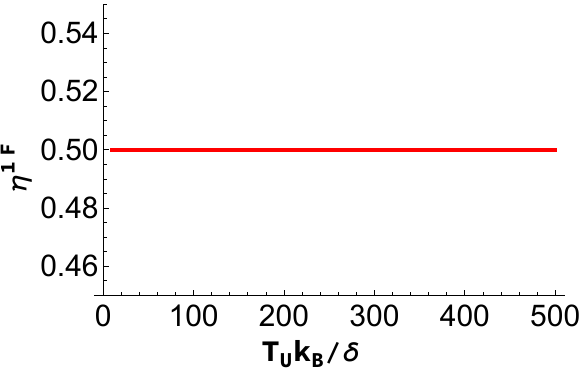}
	\caption{Behavior of the fundamental quantities describing the performance of sequential fragmented heat engines. In the left panel, we plot the maximum energy transferred to the quantum battery, ($W_M^{1F}$), that is optimized over the parameters of the engine along the vertical axis in units of $\delta$ with respect to the maximum temperature ($T_{U}$) of the hot bath that is allowed at the time of optimizing work. Along the vertical axes of the middle and right panels, the percentage of charge gained by the battery, $(C^{1F})$, and the efficiency, $(\eta^{1F})$, of the engine, are plotted for the same parameter values for which the optimal work, $W_M^{1F}$, is achieved. All the horizontal exes represent $T_U$ in units of $\delta/k_B$. In this case, the work done is a function of all the engine parameters, including the cold bath's temperature. Therefore, unlike the right panel of Fig.~\ref{fig1}, we plot the efficiency of the engine corresponding to only that temperature of the cold bath, which optimizes the work.}
	\label{fig2}
\end{figure*} 

For the Markovian approximation to be valid, we take the highest frequencies, $\omega_H^c$ or $\omega_C^c$, to be much larger than the transition frequencies, $\omega_H$ and $\omega_C$~\cite{pet}. Thus, the spectral density function reduces to $\mathcal{J}(\omega_{H/C})=\kappa \omega_{H/C}$. We take $\kappa=10^{-3}\hbar^3/\delta$. The reason behind this particular choice will be clear in Sec.~\ref{sim}.

In this setup, the three strokes of the engine, namely the heat, work, and cold strokes, occur sequentially, one after the other. In the heat stroke, $S$ is connected with the hot bath, and the evolution is governed by Eq.~\eqref{gksl} considering $\mathcal{L}(\rho)=\mathcal{L}_+^H(\rho)+\mathcal{L}_-^H(\rho)$. The interaction is switched off after $500\hbar/\delta$ time so that the system gets enough time to reach a steady state. The quantum battery to be charged is now connected to $S$ for time $t_1$. The interaction between the battery and the qutrit, which is used to charge the battery, is considered to be of the following type:
\begin{equation}
    H^{1F}_{SB}=\hbar \omega_{sb}(\tilde{a}_{12}\otimes\sigma_++\tilde{a}_{21}\otimes\sigma_-). \nonumber
\end{equation}
One can check that the commutation relation, $[H_{\text{WS}},H_{SB}^{1F}]=0$, holds for this interaction, which confirms that the total energy of the system and the quantum battery remains conserved during the work extraction process. Again, the evolution of the composite system consisting of the battery and qutrit is directed by a single parameter, $\lambda=\omega_{sb}t_1$. After time $t_1$, the working material, $S$, is connected with the cold bath, again for $500\hbar/\delta$ time. The final state of the system can be found using the GKSL master equation, Eq.~\ref{gksl}, considering $\mathcal{L}(\rho)=\mathcal{L}_+^C(\rho)+\mathcal{L}_-^C(\rho)$.

We are interested in the following characteristic quantities of a heat engine: $W_B^{1F}$, $C^{1F}$, and $\eta^{1F}$. We first numerically maximize the amount of energy that can be transferred to the quantum battery, i.e., $W_B^{1F}$, over the set of free parameters, $\{A,\lambda,T_H,T_C\}$, under the constraint $T_C<T_H$, and then determine the efficiency of the engine and PCG by the battery in that optimal set-up. For numerical simplicity, we maximize $W_B^{1F}$ by restricting ourselves within certain parameter ranges, i.e., $\{0\leq A\leq 50\delta,0\leq \lambda, 0\leq T_H,T_C\leq T_U\}$, where $T_U$ is a fixed upper bound given to the temperature of the baths.

One thing to be noted is that, since the thermal baths, be they hot or cold, are incapable of inducing transition between any pair of energy levels of $S$, unlike the previous case, the system connected to the thermal bath may not evolve to the thermal state corresponding to the given bath's temperature. Rather, it may reach another steady state, which may depend on the initial state as well as the temperature of the bath. Thus, after completion of the three strokes, the system $S$ will reach a final state, $\Bar{\rho}^{1F}_{C}$, which may not be equivalent to the initial state, $\rho_C^{1F}$. This final state, $\Bar{\rho}^{1F}_{C}$, now serves as the initial state for the second cycle. This process can be repeated again and again by taking the final state of a cycle as the initial state of the next cycle. But the quantities defining the performance of the engine, such as $W_B^{1F}$, $C^{1F}$, and $\eta^{1F}$, will be different in each round. Therefore, in the case of sequential fragmented engine, we first optimize the work done in the first round and check the final state at the end of that round, considering the parameter values for which that optimal work is reachable. That final state is found to be significantly different from the initial state. Therefore, we move on to the next round and again optimize the work done in that second cycle. We denote this optimal work done in the second cycle by $W_{M}^{1F}$ and the parameters for which this optimal value is achievable by $\{A^{1F},\lambda^{1F},T_H^{1F},T_C^{1F}\}$. The state at the end of the second cycle is obtained to be very close to the initial state of that cycle for all considered $T_U$. To be specific, the trace distance~\cite{chuang} between the initial and final states of the second round is of the order of $10^{-3}$ for $T_U<10 \delta/k_B$ and $10^{-5}$ or even smaller for $T_U>10 \delta/k_B$. Naturally, if we continue the cycles for the same parameter values, $\{A^{1F},\lambda^{1F},T_H^{1F},T_C^{1F}\}$, the work done will remain approximately the same in every round after the second cycle. Hence, we can focus on the properties of the engine in the second round.

The behavior of $W_M^{1F}$ may depend on the upper bound of the temperature, $T_U$. To understand this dependency, in Fig.~\ref{fig2}, we plot $W_{M}^{1F}/\delta$ with respect to $T_Uk_B/\delta$. It is noticeable from the figure that at first $W_{M}^{1F}/\delta$ increases with $T_Uk_B/\delta$, after which it saturates. Moreover, we find that $T_H^{1F}=T_U$, that is, similar to the case of the out-and-out sequential engine; here also, hotter baths are more advantageous. We checked for every considered $T_U$, $\lambda^{1F}=\pi/2$. For $T_U<39.16\delta/k_B$, $A^{1F}$ increases with $T_U$, and for $T_U>39.16\delta/k_B$, $A^{1F}$ becomes constant at $50\delta$, i.e., its highest allowed value specified in the numerical optimization.

In the middle panel of Fig.~\ref{fig2}, we plot the percentage of charge, $C^{1F}$, received by the quantum battery with respect to $T_U$ for the parameters that provide optimal $W_B^{1F}$. Since for $T_U>39.16\delta/k_B$, $A^{1F}$ becomes constant, in this region, $C^{1F}$ and $W_M^{1F}/\delta$ show qualitatively the same nature, that is, it increases with $T_U$ up to a certain point after which it saturates to a fixed value. Specifically, in this range, $C^{1F}=4W_M^{1F} \%$. But for $T_U<39.16\delta/k_B$, the behaviors of $C^{1F}$ and $W_M^{1F}$ are significantly different. This is because, in this regime, both the quantities, $W_M^{1F}$ and $A^{1F}$, increase with $T_U$ and both affect $C^{1F}$, where $C^{1F}$ is, on the one hand, proportional to $W_M^{1F}$ and, on the other hand, inversely proportional to $A^{1F}$. Moreover, in the right panel of the figure, we plot $\eta^{1F}$ again for the same parameter values for which $C^{1F}$ has been plotted. We would like to mention here that, in the case of this sequential fragmented engine, we have not encountered any $T_U$ for which $Q_C^{1F}$ becomes negative. 

As it can be observed from the plot, though the behaviors of $W_M^{1O}$ and $W_M^{1F}$ with respect to $T_U$ are qualitatively the same with each other, $W_M^{1F}$ is always smaller than $W_M^{1O}$ for all considered $T_U$. The same order is being followed between $C^{1F}$ and $C^{1O}$. However, the most striking difference between the fragmented sequential engine and the out-and-out sequential engine is that the efficiency in this case remains constant at $1/2$ over the temperature of the hot bath, $T_U$. Therefore, in this case, one can use the hottest bath to transfer the most amount of energy from the working substance to the quantum battery without affecting the efficiency of the engine.
 \begin{figure*}
\centering
	\includegraphics[scale=0.50]{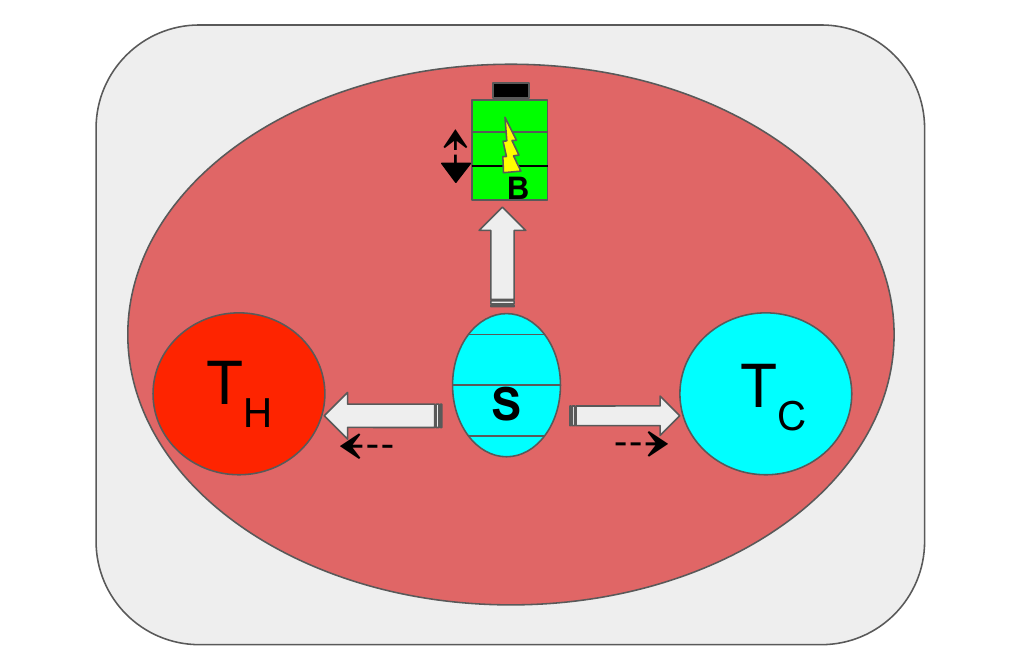}
 \hspace{.9cm}
 \includegraphics[scale=0.60]{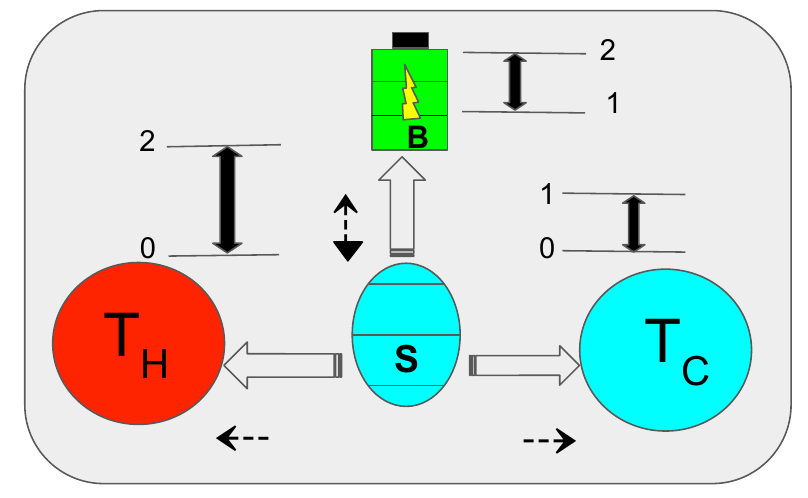}	
	\caption{Schematic diagram of simultaneous engines. The left and right diagrams illustrate interactions between the working substance with the baths and quantum battery in the first stroke for, respectively, the out-and-out and fragmented simultaneous engines. The gray solid arrows indicate each of the interactions that take place simultaneously between the constituents of the engine. Both engines undergo a second stroke, wherein the system is only connected to the cold bath to retrieve it in its initial state so that the engine can be used for another round to charge a new quantum battery. That second stroke has not been depicted in the figure. All the other considerations are the same as in Fig.~\ref{Pic1}.}
	\label{Pic2}
\end{figure*} 

\section{Simultaneous engines}
\label{sim}
Let us now combine the three distinct strokes considered in the previous section into one stroke and examine the performance of the corresponding heat engine. Precisely, unlike the sequential engine, in this case, we consider the working substance to be simultaneously interacting with the cold bath, quantum battery, and hot bath for a fixed amount of time, say $t_2$. This can be referred to as the first stroke. Whatever be the evolution time, $t_2$, unlike the sequential engine, in this case, the final state of $S$ at the end of the first stroke will not very likely be in a thermal state as it was initially. So in order to complete the cycle, we introduce a second stroke where the working substance is made to interact only with a cold bath for a long enough time for the system to get thermalized with the attached bath and reach its initial state. This makes the engine ready for the next round, where it can charge a second quantum battery and the process can be continued. In Fig.~\ref{Pic2}, we demonstrate the working methods of the sequential heat engines. 

The local energies of the working substance, baths, and the quantum battery can still be defined using Hamiltonians $H_S$, $H_X$, and $H_B$, given in Eqs.~\eqref{He}, \eqref{HX}, and \eqref{HB}, respectively. Initially, the system, $S$, is in equilibrium with the cold bath, and the battery, $B$, is in its ground state. Therefore, the composite initial state consisting of $S$ and $B$ is
\begin{equation} 
\rho_{SB}^{2p}(0)=\rho_{C}^{2p}\otimes\rho_{B}^{2p},
\end{equation}
where $ \rho_{C}^{2p}=\frac{e^\frac{-H_{S}}{k_BT_{C}}}{\Tr\left(e^\frac{-H_{S}}{k_BT_{C}}\right)}$ and $\rho_B^{2p}=\ket{0}\bra{0}$. In the first stroke, interactions are switched on between $S$ and the battery, $B$, hot bath, $H$, and cold bath, $C$. Considering the weak coupling limit, the evolution of the system-battery state, $\rho_{SB}$, in the first stroke can be determined using the GKSL master equation
 \begin{equation}
      \frac{d\rho^{2p}_{SB}}{dt}=-i[H^{2p}_{T},\rho^{2p}_{SB}]+\mathcal{L}_{H}^{2p}(\rho^{2p}_{SB})+\mathcal{L}^{2p}_{C}(\rho^{2p}_{SB}),
      \label{meq}
  \end{equation}
which is similar to Eq.~\eqref{gksl} with $H_M$ replaced by $H^{2p}_{T}=H_{S}\otimes I+ I\otimes H_{B}+H_{SB}^{2p}$, that is, the local Hamiltonian of the total system consisting of the working substance and the quantum battery. The form of the dissipative term $\mathcal{L}^{2p}_{H/C}$ depends on the properties of the interaction between the baths and $SB$.

To make the rotating wave approximation valid, we need the system relaxation time to be very large compared to $(\Delta \omega) ^{-1}$, where $\Delta \omega$ is the difference between the different angular frequencies of the bosons that get created due to the transitions occurring in $SB$ when $SB$ interacts with the baths~\cite{pet}. To slow down the evolution of the system in hand, i.e., $SB$, we consider $\kappa=10^{-3}\hbar^3/\delta$ appearing in the expression $\mathcal{J}(\omega)$. Though the rotating wave approximation is not required in the case of sequential fragmented engines, where only one transition is allowed for each of the interactions between $S$ and $X$, to compare all the engines in equal footing, we always fix $\kappa$ at $10^{-3}\hbar^3/\delta$.

The final state of $SB$ after time $t_2$ is
\begin{equation}
    \Bar{\rho}^{2p}_{B/S}=\Tr_{S/B}(\Bar{\rho}^{2p}_{SB}(t_2)).\nonumber
\end{equation}

Therefore, the second stroke starts with an initial state, $\Bar{\rho}^{2p}_{S}$, of the qutrit, $S$, and ends when $S$ reaches a thermal state, $\Bar{\rho}_S^{2p}=\rho_C^{2p}$.

Since in the first stroke, each of the two baths interacts with the system simultaneously and independently, we have two dissipative terms, $\mathcal{L}_{H}(\rho_{SB}^{2p})$ and $\mathcal{L}_{C}(\rho_{SB}^{2p})$, associated with the individual effects of hot and cold baths, respectively. The rate of heat flow to the working substance from the hot and cold baths, i.e., the heat currents, can be determined using the formula $\mathcal{J}^{2p}_H=\text{Tr}\left(H_T\mathcal{L}_{H}(\rho^{2p}_{SB})\right)$ and $\mathcal{J}^{2p}_C=\text{Tr}\left(H_T\mathcal{L}_{C}(\rho^{2p}_{SB})\right)$, respectively. The total amount of heat taken from the two baths, $Q_1^{2p}$, in the first stroke, can be calculated using the relation
\begin{equation}
   Q_1^{2p}= \int_{0~\mathcal{J}_H^{2p}>0}^{t_2}\mathcal{J}_{H}^{2p} dt+\int_{0~\mathcal{J}_C^{2p}>0}^{t_2}\mathcal{J}_{C}^{2p} dt.\nonumber\label{eq4}
\end{equation}
The total change in the battery's energy at the end of the first stroke can be used to determine the work done by the system, i.e.,
\begin{equation}
W_B^{2p}=\Tr\left[H_{B}\left(\Bar{\rho}_{B}^{2p}-{\rho}_B^{2p}\right)\right].\label{eq7}
\end{equation}
The amount of heat deposited in the cold bath in the second stroke is given by
\begin{equation*}
Q_2^{2p}= \Tr[H_{S}(\Bar{\rho}^{2p}_{S}-{\rho}^{2p}_{C})].\nonumber
\end{equation*}
$Q_2^{2p}>0$ implies heat transfer from $S$ to the cold bath and vice versa. Therefore, the total heat absorbed by the system in the two strokes is
\begin{equation}
Q_T^{2p}=Q_1^{2p}-Q_2^{2p}\theta\left(-Q_2^{2p}\right). \label{eq9}
\end{equation}
Once $Q_T^{2p}$ and $W_B^{2p}$ are in hand, $\eta^{2p}$ and $C^{2p}$ can be easily determined.

These simultaneous engines can also operate in two ways, which, like the sequential engine, can be referred to as the out-and-out engines and the fragmented engines [see Fig.~\ref{Pic2} for deeper understanding] and can be represented by $p=O$ and $p=F$, respectively. Let us now discuss each of these two types of engines separately.

\subsection{Out-and-out simultaneous engine}
\begin{figure*}
\centering
	\includegraphics[scale=0.59]{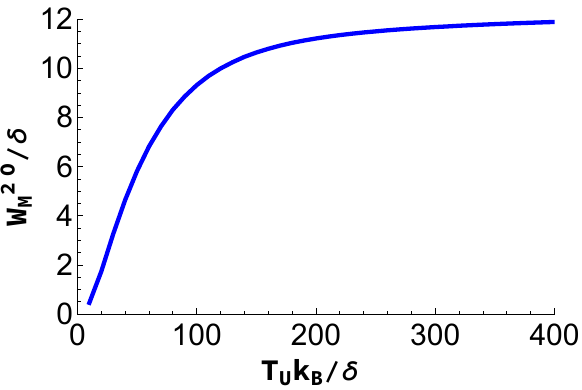}
	\includegraphics[scale=0.59]{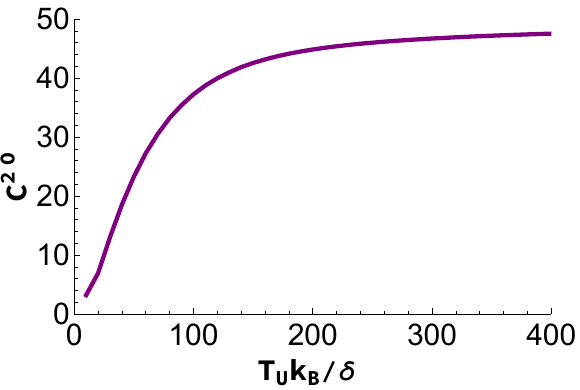}
	\includegraphics[scale=0.59]{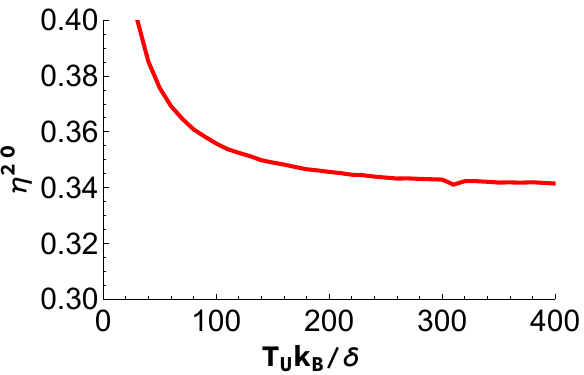}
	\caption{Depiction of the performance of the simultaneous out-and-out engine. The vertical axes of the left, middle, and right panels present the energy transferred to the quantum battery, $W_M^{2O}$, PCG by the battery, $C^{2O}$, and efficiency of the engine, $\eta^{2O}$. The quantities shown in the vertical axes of all the panels are plotted for those parameters of the engines for which the battery can receive the most energy. The horizontal axes depict $T_U$, the upper bound on the temperature of the hot bath restricted at the time of numerically optimizing the charge gained by the battery. All the axes are dimensionless. 
 }
	\label{fig3}
\end{figure*} 

The out-and-out simultaneous engine is defined in a similar way as in the case of the sequential engine, i.e., in this engine, the baths, when connected to the working substance, are considered to be capable of inducing a population transition between any two energy levels of the system, $S$. Hence, the interaction Hamiltonian is again given by
\begin{equation}
\label{2}
\begin{split}
H^{2O}_{SX}&=\int_{0}^{\omega_{X}^c}h(\omega)\Big[\tilde{a}_{01}\left(a_{\omega}^X\right)^{\dagger } +\tilde{a}_{10}a_{\omega}^X+\tilde{a}_{02}\left(a_{\omega}^X\right)^{\dagger }  
\\&+\tilde{a}_{20}a_{\omega}^{X}
+\tilde{a}_{12}\left(a_{\omega}^X\right)^{\dagger } +\tilde{a}_{21}\left(a_{\omega}^X\right)^{\dagger }\Big]d\omega.
\end{split}
\end{equation}
The interaction between $S$ and $B$ is considered to be of the following form
\begin{equation}
\label{Hsb}
H^{2O}_{{S{B}}}=\hbar\omega_{sb}(J_{+}\otimes\sigma_{-}+J_{-}\otimes\sigma_{+}),
\end{equation}
which is the same as the case of an out-and-out sequential engine. The joint initial state of $S$ and $B$ is $\rho_{SB}^{2O}=\rho_{C}^{2O}\otimes\rho_{B}^{2O}$ where $\rho_C^{2O}$ and $\rho_B^{2O}$ are, respectively, the thermal state of $S$ with temperature $T_C$ and the ground state of the quantum battery. The evolution of the system-battery state when the system is involved in interaction with the hot and cold baths is governed by Eq.~\eqref{meq} with $p=O$. The exact mathematical form of $\mathcal{L}^{2O}_{H/C}$ is provided in Appendix~\ref{A1}. 

The amount of energy transferred to a quantum battery and the total amount of heat taken by $S$ from the baths within a cycle can be numerically determined using Eqs.~\eqref{eq7} and \eqref{eq9}, respectively, which can then be used to quantify $C^{2O}$ and $\eta^{2O}$. To understand the performance of the engine in an optimal situation, we maximize the total work output, $W_B^{2O}$, over the set of free parameters, $\{A,T_H,T_C,\omega_{sb},t_2\}$. The optimization is done numerically within the range $\{0\leq A\leq 50\delta,0\leq \omega_{sb}\leq 25\delta/\hbar, 0\leq T_H\leq T_U,0\leq T_C\leq T_U\, 0\leq t_2 \leq 10\hbar/\delta\}$. We have considered $\omega_{sb}$ to be less than $A/2$. The reason behind this consideration can be found in Appendix~\ref{A1}. We denote the parameters for which the optimal work, $W_M^{2O}$, can be achieved by $\{A^{2O},T_H^{2O},T_C^{2O},\omega_{sb}^{2O},t_2^{2O}\}$. We find that for smaller $T_U$, $A^{2O}$ increases with $T_U$; for $T_U\geq 30 \delta/k_B$, $A^{2O}$ becomes constant at $50\delta$. $t_2^{1O}$ and $T_H^{1O}$ are also found to be equal to the highest allowed value in the optimization algorithm, i.e., 10$\hbar/\delta$ and $T_U$. Moreover, we find $T_C^{1O}\approx T_H^{1O}=T_U$ for all $T_U$. We witness that for the optimal set of parameters, $\{A^{2O},T_H^{2O},T_C^{2O},\omega_{sb}^{2O},t_2^{2O}\}$, the working substance absorbs heat from both baths. One can notice from Eq.~\eqref{2} that the interaction Hamiltonian is the same for both hot and cold baths. The only difference is in the temperatures of the baths, which, in the case of the most profitable engine with respect to the energy transferred to the battery, becomes almost equal. Thus, we can conclude that in the case of these optimal simultaneous out-and-out engines, both baths become equivalent to each other. It is important to note here that in the case of this optimal engine, we never encountered a situation where $S$ absorbs heat in the second stroke.

The natures of $W_B^{2O}$ or $W_M^{2O}$ (left panel), $C^{2O}$ (middle panel), and $\eta^{2O}$ (right panel) are demonstrated in Fig.~\ref{fig3} with respect to $T_U$ considering the optimal set of parameters, $\{A^{2O},T_H^{2O},T_C^{2O},\omega_{sb}^{2O},t_2^{2O}\}$. For this set of parameter values, $W_B^{2O}$ and $W_M^{2O}$ are the same. One can notice from the figure that both $W_B^{2O}$ and $C^{2O}$ monotonically increase with $T_U$ saturating to a constant value for large $T_U$. On the other hand, the efficiency shows exactly the opposite nature, that is, $\eta^{2O}$ at first decreases with $T_U$ finally acquiring a steady value. 

 \subsection{Fragmented simultaneous engine}
 \begin{figure*}
\centering
	\includegraphics[scale=0.59]{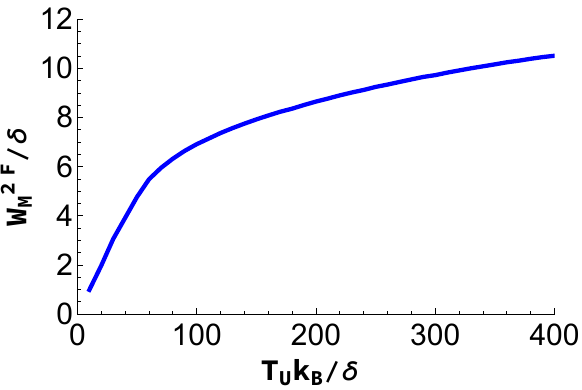}
	\includegraphics[scale=0.59]{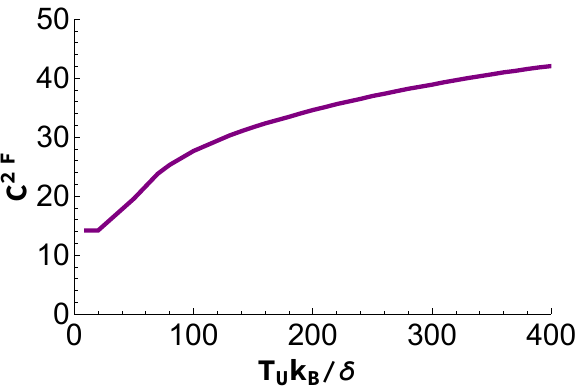}
	\includegraphics[scale=0.59]{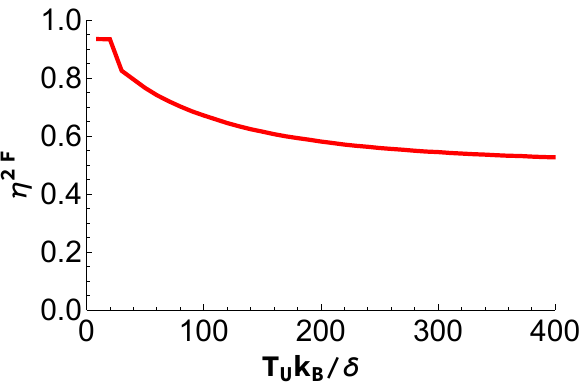}
	\caption{Illustration of the performance of a simultaneous fragmented engine through its characteristic quantities We optimize $W_B^{2F}$ over the parameters describing the engine and show the behavior of that maximum work done, $W_M^{2F}$, in the left panel along with the corresponding nature of $C^{2F}$ and $\eta^{2F}$, which are presented, respectively, in the middle and right panels, for the same parameters for which $W_M^{2F}$ is plotted. $T_U$ presented along the horizontal axes of all of the plots denotes the upper bound of temperature given to the hot bath at the time of work optimization. 
 }
	\label{fig4}
\end{figure*} 
In the fragmented model of simultaneous engines, the baths are connected with $S$ in such a way that the baths can make a population transition between any two particular energy levels of $S$. The form of the interaction Hamiltonians, $H_{SH}$, $H_{SC}$, and $H_{SB}$, in this case, are considered to be of the following forms:
\begin{eqnarray*}
H_{SH}^{2F}&=\int_{0}^{\omega_{X}^c}h(\omega)\Big[\tilde{a}_{02}\left(a_{\omega}^X\right)^{\dagger } +\tilde{a}_{20}a_{\omega}^X\Big],
\\
H_{SC}^{2F}&=\int_{0}^{\omega_{X}^c}h(\omega)\Big[\tilde{a}_{01}\left(a_{\omega}^X\right)^{\dagger } +\tilde{a}_{10}a_{\omega}^X\Big],~\text{and}\nonumber\\
H^{2F}_{{S{B}}}&=\hbar\omega_{sb}(\tilde{a}_{21}\otimes\sigma_{-}+\tilde{a}_{12}\otimes\sigma_{+}).
\end{eqnarray*}
These are exactly the same interaction Hamiltonians that have been considered in sequential fragmented engines.

Let us now discuss the evolution of the engine. The initial system-battery state is $\rho_{SB}^{2F}=\rho_{C}^{2F}\otimes\rho_{B}^{2F}$ where $\rho_{C}^{2F}$ and $\rho_{B}^{2F}$ respectively denote the thermal state of $S$ with temperature $T_C$ and the ground state of $B$. Like previously, here also the joint evolution of $S$ and $B$ in the first stroke can be obtained using the GKSL master equation given in Eq.~\eqref{meq}. Expressions of $\mathcal{L}_{H}^{2F}$ and $\mathcal{L}_{C}^{2F}$ associated with the hot and cold baths are presented in Appendix~\ref{A2}. We have introduced a second stroke in the case of simultaneous engines, where the working substance, $S$, interacts with the cold bath, $C$, so that we can thermalize $S$ and bring it to its initial state. However the fragmented type of interactions being incapable of making transitions between any two energy levels $S$ does not have a unique steady state. Therefore, if we consider a fragmented-type interaction between $S$ and $C$ again in the second stroke, whatever be the time interval of the second stroke, it may not always be possible to get a thermal state at the end of the stroke. Thus, for the second stroke, we introduce a different kind of bath that is still bosonic and has a temperature of $T_C$, but interacts with $S$ in an out-and-out way, making a transition possible between any two energy levels of $S$. Hence, because of this interaction between the bath and the qutrit, at the end of the second stroke, we retrieve the initial state. 
  
The total work done, $W_B^{2F}$, by $S$ in charging the quantum battery and the total amount of heat extracted from the baths, $Q_T^{2F}$, can be determined using the expressions given in \eqref{eq7} and \eqref{eq9}. We again optimize the work output, $W_B^{2F}$, over the set of parameters $\{0\leq A\leq 50\delta,0\leq \omega_{sb} \leq 25\delta/\hbar , 0\leq T_H\leq T_U,0\leq T_C\leq T_U, 0\leq t_2 \leq 10\hbar/\delta\}$. We have imposed the constraint that $\omega_{sb}<\frac{A}{2}$. The justification behind this consideration can be found in Appendix~\ref{A2}.

We find that in this case also $T_H^{2F}=T_U$ and $t_2^{2F}$ is close to 10$\hbar/\delta$ where $\{ A^{2F}, \omega_{sb}^{2F},T_H^{2F},T_C^{2F},t_2^{2F}\}$ denotes the set of parameters for which the optimal work, $W_B^{2F}$, can be achieved. $A^{2F}$ is found to be increasing with $T_U$ for $T_U<40\delta/k_B$, above which $A^{2F}$ saturates at 50$\delta$. The hot bath's temperature, $T_H^{2F}$, in this case, is found to be significantly larger than $T_C^{2F}$ but again equal to $T_U$. Additionally, the working substance is found to only absorb heat from the hot bath and not from the cold bath in the first stroke. That is, in the first stroke, the heat bath behaves as a heat source while the cold bath plays the role of a heat sink. The working substance, $S$, extracts heat from the hot bath and feeds it to the cold bath with a side channel to the quantum battery due to the interaction between the battery and $S$. We find that, in the case of this simultaneous fragmented engine with parameters $\{ A^{2F}, \omega_{sb}^{2F},T_H^{2F},T_C^{2F},t_2^{2F}\}$, though $S$ does not absorb heat from the cold bath in the first stroke, it consumes heat in the second stroke. The reason behind absorbing heat in the second stroke may be the same as discussed in the case of the sequential out-and-out engines in Sec.~\ref{ok}.

In the left panel of Fig.~\ref{fig4}, we show the dependency of the optimal work $W_M^{2F}$ on $T_U$. In addition to $W_M^{2F}$, we also plot $C^{2F}$ and $\eta^{2F}$ in the middle and right panels of the same figure with respect to $T_U$ for the parameters that provide optimal work. One can observe from the figures that in this case also higher temperature baths cause higher energy transfer to the quantum battery, though the efficiency is lesser.

\label{ok2}
 \section{Comparison between the engines}
 \label{seclast}

  \begin{figure}
\centering
	\includegraphics[scale=0.6]{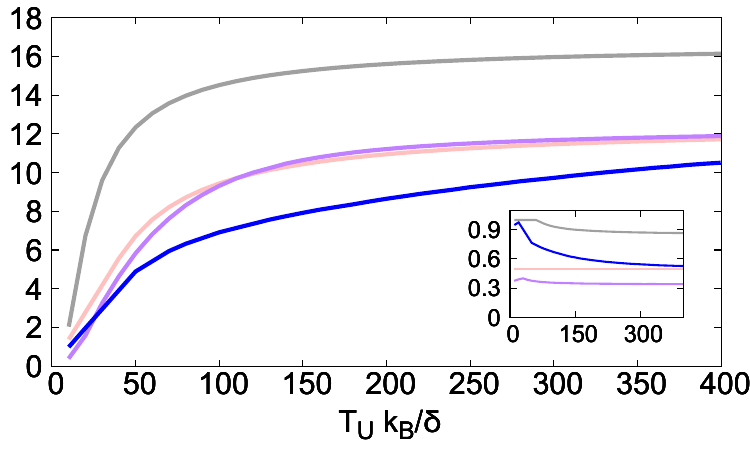}
	\caption{Difference in the performance of the heat engines. We plot the variation of the maximum work done by the sequential out-and-out (dark-gray curve), sequential fragmented (pink curve), simultaneous out-and-out (purple curve), and simultaneous fragmented (blue curve) engines in units of $\delta$ with respect to $T_U$, i.e., the highest temperature the baths are allowed to have. $T_U$ is plotted in the unit of $k_B/\delta$. These are the same curves that have been individually shown in the left panels of Figs.~\ref{fig1}, \ref{fig2}, \ref{fig3}, and \ref{fig4}, respectively. The inset shows the behavior of the corresponding efficiencies of these engines with respect to the same quantity, $T_U k_B/\delta$, when the work done is optimal. The dark-gray, pink, purple, and blue curves of the inset showing the nature of the efficiencies of the engines are the same as the yellow curve of the right panel of Fig.~\ref{fig1} and the red curves of the right panels of Figs.~\ref{fig2}, \ref{fig3}, and \ref{fig4}, respectively. All the axes are dimensionless.}
	\label{figLast}
\end{figure}
 In this work, we have examined the performance of four different engines: sequential out-and-out, sequential fragmented, simultaneous out-and-out, and simultaneous fragmented. In Fig.~\ref{figLast}, we plot the optimal work done by the four types of engines with respect to $T_U k_B/\delta$, where $T_U$ represents the highest temperature the baths are allowed to have during optimization. From the figure, it is evident that the qualitative behavior of the optimal work done, $W_M^P$, by the engines with respect to $T_U$ is the same for all considered heat engines. Specifically, $W_M^P$, at first, monotonically increases with $T_U$ and finally attains an almost constant value. Nevertheless, the quantitative nature of the work done significantly varies from engine to engine. Analyzing the behavior of the optimal work done by the engines, one can easily conclude that the engines follow a specific order among themselves [see Fig.~\ref{figLast}]. Precisely, sequential out-and-out $\stackrel{W}{\succ}$ sequential fragmented $\stackrel{W}{\approx}$ simultaneous out-and-out $\stackrel{W}{\succ}$ simultaneous fragmented, where $\stackrel{W}{\succ}$ and $\stackrel{W}{\approx}$ denote respectively greater or almost equivalent utility with respect to work output. However, as one can see from the inset of Fig.~\ref{figLast}, the ranking gets changed if one focuses on the efficiency of the engines operating optimally to provide maximum work. In a similar manner, greater utility with respect to efficiency can be represented using the symbol $\stackrel{\eta}{\succ}$. The order followed by the engines with respect to that efficiency is the following: simultaneous fragmented $\stackrel{\eta}{\succ}$ sequential fragmented $\stackrel{\eta}{\succ}$ simultaneous out-and-out. The efficiency of the sequential out-and-out engine can be varied by changing the temperature of the cold bath while keeping the work done fixed at the maximum value. Thus, the sequential out-and-out engine does not occupy a particular position in the ordered list with respect to efficiency. But its efficiency can be easily increased by increasing the cold bath's temperatures and being made equal to unity for a high range of temperature of the hot bath, which makes the engine the most efficient one among the lot. 

\section{Conclusion}
\label{Con} 

In this article, we explored the features of sequential and simultaneous engines. Sequential engines under consideration involve three consecutive strokes, namely the heat, work, and cold strokes, during which the working substance interacts successively with a hot bath, a cold bath, and a quantum battery, respectively. To model the simultaneous engines, we merged these three distinct strokes into one, allowing the working substance to simultaneously interact with all the baths and the quantum battery. The engine's primary objective is to extract heat from the available baths and transfer most of its energy to the battery. While we expected the working medium to absorb heat from the hotter bath and release heat to the colder bath, we encountered exceptions in our analysis.

Furthermore, we categorized both the sequential and simultaneous engines into two parts based on the interactions between the working substance and the baths or the battery. These categories are the out-and-out engines, where the interaction can induce a transition between any two energy levels of the working substance, and the fragmented engines, where each of the baths and quantum battery can induce a transition between a specific pair of energy levels of the working substance.

To assess the performance of the four-engine types, namely sequential out-and-out, sequential fragmented, simultaneous out-and-out, and simultaneous fragmented, we analyzed several figures of merit, that is, the amount of energy transferred to the battery, the percentage of energy gained by the battery, and the engine's efficiency. We maximized the charge gained by the quantum battery while keeping the upper bound on the temperature of the hot bath fixed. We determined that the work achieved its maximum value only when the temperature of the hot bath reached its upper limit. We presented the behavior of the optimal work done, the corresponding percentage of charge gained by the quantum battery, and the efficiency of the engine with respect to the hot bath's temperature. By ``corresponding'', we mean the same parameter values for which maximum energy can be transferred to the battery.

Across all four engine types, these characteristics displayed qualitatively similar behavior, except for the efficiency of the sequential fragmented engine, which remained constant regardless of the hot bath's temperature. Except for the sequential fragmented engines, the work done and the engine efficiency exhibited a complementary relationship with respect to the hot bath's temperature: work increased as temperature rose, while efficiency decreased.

We found that the sequential out-and-out engines were the most profitable engines, boasting both the highest efficiency and the capacity to transfer the largest amount of energy to the quantum batteries when the cold baths' temperature is sufficiently large. The other three engines did not exhibit a specific performance order. For instance, when comparing work output, the simultaneous out-and-out engines seemed more advantageous than the simultaneous fragmented engines, but in terms of efficiency, the simultaneous fragmented engines outperformed the other two. From Figs.~\ref{fig3} and \ref{fig4}, it is evident that the simultaneous out-ant-out engines provide greater work while the fragmented simultaneous engines achieve higher efficiency corresponding to the parameters for which each of the engines transfers maximum energy to the attached quantum battery in the first stroke.

\section{Acknowledgement}
We acknowledge partial support from the Department of Science and Technology, Government of India through the QuEST grant (grant number DST/ICPS/QUST/Theme3/2019/120).
\appendix
  \section{Explicit forms of the Lindblad operators of out-and-out simultaneous engine} 
  \label{A1}
 The eigenvectors, $\{\ket{\boldsymbol{i}}\}$, and the corresponding eigenvalues, $\{E_{i}\}$, of $H_{T}^{2O}$ are given by
       \begin{itemize}
       \item$\ket{\textbf{0}}=\ket{\tilde{0}0}, E_{0}=-\frac{3A}{4}$
        \item$\ket{\textbf{1}}=\frac{\ket{\tilde{1}0}-\ket{\tilde{0}1}}{\sqrt{2}},E_{1}=-\frac{A+4\hbar\omega_{sb}}{4}$
    \item$\ket{\textbf{2}}=\frac{\ket{\tilde{1}0}+\ket{\tilde{0}1}}{\sqrt{2}},E_{2}=\frac{-A+4\hbar\omega_{sb}}{4}$
          \item$\ket{\textbf{3}}=\frac{\ket{\tilde{1}1}-\ket{\tilde{2}0}}{\sqrt{2}},E_{3}=\frac{A-4\hbar\omega_{sb}}{4}$
    \item$\ket{\textbf{4}}=\frac{\ket{\tilde{1}1}+\ket{\tilde{2}0}}{\sqrt{2}},,E_{4}=\frac{A+4\hbar\omega_{sb}}{4}$
    \item$\ket{\textbf{5}}=\ket{\tilde{2}1},E_{5}=\frac{3A}{4}$
       \end{itemize}
To maintain the order, that is, $E_i\leq E_{i+1}$ for all $i$ between 0 to 5, we consider $\omega_{sb}\leq\frac{A}{2\hbar}$.

Since the baths are connected only with $S$ and not with $B$, transitions will only occur between the energy levels of $S$ specified by $H_S$. The Lindblad operators defining  possible transitions and the corresponding transition gaps are given by 
  \begin{itemize}
      \item $A_{1}=(\ket{\textbf{0}}\bra{\textbf{2}}+\ket{\textbf{3}}\bra{\textbf{5}})/\sqrt{2},~ \omega_{1}=(A+2\hbar\omega_{sb})/2$
      \item $A_{2}=(\ket{\textbf{0}}\bra{\textbf{1}}+\ket{\textbf{4}}\bra{\textbf{5}})/\sqrt{2},~\omega_{2}=(A-2\hbar\omega_{sb})/2$
      \item $A_{3}=(\ket{\textbf{0}}\bra{\textbf{4}}-\ket{\textbf{1}}\bra{\textbf{5}})/\sqrt{2},~\omega_{3}=A+\hbar\omega_{sb}$
      \item $A_{4}=(\ket{\textbf{2}}\bra{\textbf{5}}-\ket{\textbf{0}}\bra{\textbf{3}})/\sqrt{2},~\omega_{4}=A-\hbar\omega_{sb}$
      \item $A_{5}=\ket{\textbf{2}}\bra{\textbf{4}}- \ket{\textbf{1}}\bra{\textbf{3}},~\omega_{5}=A/2$.
  \end{itemize} 
 We can again consider $\mathcal{L}^X(\rho)=\mathcal{L}_+^X(\rho)+\mathcal{L}_-^X(\rho)$ where 
\begin{eqnarray*}
\label{l2}
 \mathcal{L}_+^X(\rho)=
 \sum_{i=7}^2\mathcal {J}(\omega_{i})(1+n(\omega_{i},T_X))\Big[A_{i}(\omega_{i})\rho A_{i}^\dagger(\omega_{i})\\
-\frac{1}{2}\{\rho,A_{i}^\dagger(\omega_{i}) A_{i}(\omega_{i})\}\Big]\\
    \mathcal{L}_-^X(\rho)=\sum_{i=1}^7\mathcal{J}(\omega_{i})n(\omega_{i},T_X)\Big[A_{i}(\omega_{i})^\dagger \rho A_{i}(\omega_{i})\\
- \frac{1}{2}\{\rho,A_{i}(\omega_{i}) A_{i}^\dagger(\omega_{i})\}\Big]
\end{eqnarray*}
The only difference between $\mathcal{L}_{H}(\rho_{SB})$ and $\mathcal{L}_{C}(\rho_{SB})$ is in $n(\omega_{i},T_X)$, which depends on the temperature, $T_X$, of the respective baths.

 \section{Explicit forms of the Lindblad operators of fragmented simultaneous engine}
 \label{A2}
The eigenvalues, $\{E_i\}$, and eigenvectors, $\{\ket{\boldsymbol{i}}\}$, of the total Hamiltonian, $H_T^{2F}$, are given by
       \begin{itemize}
       \item $\ket{\textbf{0}}=\ket{\tilde{0}0}, E_{1}=-\frac{3A}{4}$
        \item $\ket{\textbf{1}}=\ket{\tilde{1}0},E_{2}=-\frac{A}{4}$
         \item $\ket{\textbf{2}}=\ket{\tilde{0}1},E_{3}=-\frac{A}{4}$
          \item $\ket{\textbf{3}}=\frac{\ket{\tilde{1}1}-\ket{\tilde{2}0}}{\sqrt{2}},E_{4}=\frac{A-4\hbar\omega_{sb}}{4}$
           \item $\ket{\textbf{4}}=\frac{\ket{\tilde{1}1}+\ket{\tilde{2}0}}{\sqrt{2}},E_{5}=\frac{A+4\hbar\omega_{sb}}{4}$
            \item $\ket{\textbf{5}}=\ket{\tilde{2}1},E_{6}=\frac{3A}{4}$.
       \end{itemize}
  We consider $E_i\leq E_{i+1}$ for all $i=$ between0 to 5 which implies $\omega_{sb}<\frac{A}{2}$.
  In the case of a simultaneous fragmented heat engine, the hot bath is connected to the qutrit in such a way that it governs the transition between the ground and the second excited levels of the attached qutrit. Therefore the corresponding Lindblad operators and transition energies can be written as
  \begin{itemize}
  \item $A_{1}^{H}=-\ket{\boldsymbol{0}}\bra{\boldsymbol{3}},\omega_{1}^H=A-\hbar\omega_{sb}$
  \item $A_{2}^{H}=\ket{\boldsymbol{0}}\bra{\boldsymbol{4}},\omega_{2}^H=A+\hbar\omega_{sb}$
   \item $A_{3}^{H}=\ket{\boldsymbol{2}}\bra{\boldsymbol{5}},\omega_{3}^H=A$.
   \end{itemize}
   The cold bath can make a transition between the first excited and ground state of the qutrit. The Lindblad operators and the corresponding transition energies in this case are
   \begin{itemize}
  \item $A_{1}^{C}=\ket{\boldsymbol{0}}\bra{\boldsymbol{1}},\omega_{1}^C=\frac{A}{2}$
  \item $A_{2}^{C}=\ket{\boldsymbol{2}}\bra{\boldsymbol{3}},\omega_{2}^C=\frac{A}{2}-\hbar\omega_{sb}$
   \item $A_{3}^{C}=\ket{\boldsymbol{2}}\bra{\boldsymbol{4}},\omega_{3}^C=\frac{A}{2}+\hbar\omega_{sb}$.\\
   \end{itemize}
  The expressions of $\mathcal{L}_H(\rho)$ and $\mathcal{L}_C(\rho)$ are given by
\begin{widetext}
\begin{eqnarray*}
     \mathcal{L}^X(\rho)=
 \sum_{i=1}^3\mathcal {J}(\omega_{i})(1+n(\omega_{i},T_X))\Big[A^X_{i}(\omega_{i})\rho \left(A^X_{i}(\omega_{i})\right)^\dagger-\frac{1}{2}\{\rho,\left(A_{i}^X(\omega_{i})\right)^\dagger A^X_{i}(\omega_{i})\}\Big]+\sum_{i=1}^3\mathcal{J}(\omega_{i})n(\omega_{i},T_X)\nonumber\\\Big[\left(A^X_{i}(\omega_{i})\right)^\dagger \rho A^X_{i}(\omega_{i})
- \frac{1}{2}\{\rho,A^X_{i}(\omega_{i}) \left(A^X_{i}(\omega_{i})\right)^\dagger\}\Big],
\end{eqnarray*}
\end{widetext}
where $X=H$ and $C$ represent the hot and cold baths, respectively.

\bibliography{3SHEref}
\end{document}